\documentclass[12pt]{article}

\catcode`\@=11
\def\@citex[#1]#2{%
\if@filesw \immediate \write \@auxout {\string \citation {#2}}\fi
\@tempcntb\m@ne \let\@h@ld\relax \def\@citea{}%
\@cite{%
  \@for \@citeb:=#2\do {%
    \@ifundefined {b@\@citeb}%
      {\@h@ld\@citea\@tempcntb\m@ne{\bf ?}%
      \@warning {Citation `\@citeb ' on page \thepage \space undefined}}%
      {\@tempcnta\@tempcntb \advance\@tempcnta\@ne%
      \@tempcntb\number\csname b@\@citeb \endcsname \relax%
      \ifnum\@tempcnta=\@tempcntb 
        \ifx\@h@ld\relax%
          \edef \@h@ld{\@citea\csname b@\@citeb\endcsname}%
        \else%
          \edef\@h@ld{\ifmmode{-}\else--\fi\csname b@\@citeb\endcsname}%
        \fi%
      \else
        \@h@ld\@citea\csname b@\@citeb \endcsname%
        \let\@h@ld\relax%
      \fi}%
    \def\@citea{,\penalty\@highpenalty\,}%
  }\@h@ld
}{#1}}

\def\@citeb#1#2{{[#1]\if@tempswa , #2\fi}}
\def\@citeu#1#2{{$^{#1}$\if@tempswa , #2\fi }}
\def\@citep#1#2{{#1\if@tempswa , #2\fi}}

\def\bcites{         
        \catcode`\@=11
        \let\@cite=\@citeb
        \catcode`\@=12
}

\def\upcites{         
        \catcode`\@=11
        \let\@cite=\@citeu
        \catcode`\@=12
}

\def\plaincites{      
        \catcode`\@=11
        \let\@cite=\@citep
        \catcode`\@=12
}

\newcount\hour
\newcount\minute
\newtoks\amorpm
\hour=\time\divide\hour by 60
\minute=\time{\multiply\hour by 60 \global\advance\minute by-\hour}
\edef\standardtime{{\ifnum\hour<12 \global\amorpm={am}%
        \else\global\amorpm={pm}\advance\hour by-12 \fi
        \ifnum\hour=0 \hour=12 \fi
        \number\hour:\ifnum\minute<10 0\fi\number\minute\the\amorpm}}
\edef\militarytime{\number\hour:\ifnum\minute<10 0\fi\number\minute}

\def\draftlabel#1{{\@bsphack\if@filesw {\let\thepage\relax
   \xdef\@gtempa{\write\@auxout{\string
      \newlabel{#1}{{\@currentlabel}{\thepage}}}}}\@gtempa
   \if@nobreak \ifvmode\nobreak\fi\fi\fi\@esphack}
        \gdef\@eqnlabel{#1}}
\def\@eqnlabel{}
\def\@vacuum{}
\def\marginnote#1{}
\def\draftmarginnote#1{\marginpar{\raggedright\scriptsize\tt#1}}
\def\draft{
        \pagestyle{plain}
        \overfullrule=2pt
        \oddsidemargin -.5truein
        \def\@oddhead{\sl \phantom{\today\quad\militarytime} \hfil
        \smash{\Large\sl DRAFT} \hfil \today\quad\militarytime}
        \let\@evenhead\@oddhead
        \let\label=\draftlabel
        \let\marginnote=\draftmarginnote
        \def\ps@empty{\let\@mkboth\@gobbletwo
        \def\@oddfoot{\hfil \smash{\Large\sl DRAFT} \hfil}
        \let\@evenfoot\@oddhead}
        \def\@eqnnum{(\theequation)\rlap{\kern\marginparsep\tt\@eqnlabel}%
        \global\let\@eqnlabel\@vacuum}  }

\def\blackfonts{
        \font\blackboard=msbm10 scaled\magstep1
        \font\blackboards=msbm8
        \font\blackboardss=msbm6
}

\def\prep{         
        \catcode`\@=11
        \input art10.sty
        \catcode`\@=12
        
        \let\small\null
        \def\blackfonts{
                \font\blackboard=msbm10
                \font\blackboards=msbm7
                \font\blackboardss=msbm5
        }
        \let\sl\it
        \twocolumn
        \sloppy
        \voffset=-2.54truecm
        \hoffset=-2.54truecm
        \flushbottom
        \parindent 1em
        \leftmargini 2em
        \leftmarginv .5em
        \leftmarginvi .5em
        \marginparwidth 48pt
        \marginparsep 10pt
        \setlength{\columnsep}{2truecm}
        \setlength{\textwidth}{25.4truecm}
        \setlength{\textheight}{17truecm}
        \baselineskip=16pt
        \oddsidemargin .18truein
        \evensidemargin .17truein
}

\def\eqalign#1{\null\,\vcenter{\openup\jot\m@th
  \ialign{\strut\hfil$\displaystyle{##}$&$\displaystyle{{}##}$\hfil
      \crcr#1\crcr}}\,}
\def\eqalignno#1{\displ@y \tabskip\centering
  \halign to\displaywidth{\hfil$\@lign\displaystyle{##}$\tabskip\z@skip
    &$\@lign\displaystyle{{}##}$\hfil\tabskip\centering
    &\llap{$\@lign##$}\tabskip\z@skip\crcr
    #1\crcr}}

\def\section{\@startsection {section}{1}{\z@}{3.ex plus 1ex minus
 .2ex}{2.ex plus .2ex}{\large\bf}}
\def\subsection{\@startsection{subsection}{2}{\z@}{2.75ex plus 1ex minus
 .2ex}{1.5ex plus .2ex}{\bf}}        

\def\appendix{{\newpage\section*{Appendix}}\let\appendix\section%
        {\setcounter{section}{0}
        \gdef\thesection{\Alph{section}}}\section}

\def\abstract{\if@twocolumn
\section*{Abstract}
\else 
\begin{center}
{\bf Abstract\vspace{-.5em}\vspace{0pt}}
\end{center}
\quotation
\fi}

\catcode`\@=12

\def\d{\partial}

\def\sqr#1#2{{\vcenter{\vbox{\hrule height.#2pt\hbox{\vrule width.#2pt 
height#1pt \kern#1pt \vrule width.#2pt}\hrule height.#2pt}}}}

\def\=d{\,{\buildrel\rm def\over =}\,}

\def\N{\hbox{\bbf N}}

\def\i3p{\p32\int d^3p}

\def\As{A\hbox to 1pt{\hss /}}
\def\np4{\int d^4p_1\cdots d^4p_{n-1}\, }

\def\tr{{\rm tr}\, }

\def\nx4{\int d^4x_1\ldots d^4x_n\, }

\def\kon#1#2{\vbox{\halign{##&&##\cr
\lower4pt\hbox{$\scriptscriptstyle\vert$}\hrulefill &
\hrulefill\lower4pt\hbox{$\scriptscriptstyle\vert$}\cr $#1$&
$#2$\cr}}}

\def\konv#1#2#3{\hbox{\vrule height12pt depth-1pt}
\vbox{\hrule height12pt width#1cm depth-11.6pt}
\hbox{\vrule height6.5pt depth-0.5pt}
\vbox{\hrule height11pt width#2cm depth-10.6pt\kern5pt
      \hrule height6.5pt width#2cm depth-6.1pt}
\hbox{\vrule height12pt depth-1pt}
\vbox{\hrule height6.5pt width#3cm depth-6.1pt}
\hbox{\vrule height6.5pt depth-0.5pt}}
\def\konu#1#2#3{\hbox{\vrule height12pt depth-1pt}
\vbox{\hrule height1pt width#1cm depth-0.6pt}
\hbox{\vrule height12pt depth-6.5pt}
\vbox{\hrule height6pt width#2cm depth-5.6pt\kern5pt
      \hrule height1pt width#2cm depth-0.6pt}
\hbox{\vrule height12pt depth-6.5pt}
\vbox{\hrule height1pt width#3cm depth-0.6pt}
\hbox{\vrule height12pt depth-1pt}}

\def\konw#1#2#3{\hbox{\vrule height12pt depth-1pt}
\vbox{\hrule height12pt width#1cm depth-11.6pt}
\hbox{\vrule height6.5pt depth-0.5pt}
\vbox{\hrule height12pt width#2cm depth-11.6pt \kern5pt
      \hrule height6.5pt width#2cm depth-6.1pt}
\hbox{\vrule height6.5pt depth-0.5pt}
\vbox{\hrule height12pt width#3cm depth-11.6pt}
\hbox{\vrule height12pt depth-1pt}}

\def\i{{\rm int}}

\def\m3{{\mu_1\mu_2\mu_3}}

\def\p{{(+)}}

\def\be{\begin{equation}}       \def\eq{\begin{equation}}
\def\ee{\end{equation}}         \def\eqe{\end{equation}}

\def\bea{\begin{eqnarray}}      \def\eqa{\begin{eqnarray}}
\def\ena{\end{eqnarray}}        \def\eea{\end{eqnarray}}
                                \def\eqae{\end{eqnarray}}

\def\ba{\begin{array}}
\def\ea{\end{array}}
\def\unit{1 \hskip-.3em \raise2pt\hbox{$ \scriptstyle |$ } }

\def\d{\delta}

\def\i{\iota}

\def\m{\mu}
\def\n{\nu}
  
\def\p{\pi}                
\def\t{\tau}

\def\D{\Delta}

\def\G{\Gamma}
\def\J{\Psi}

\def\bop#1{\setbox0=\hbox{$#1M$}\mkern1.5mu
        \vbox{\hrule height0pt depth.04\ht0
        \hbox{\vrule width.04\ht0 height.9\ht0 \kern.9\ht0
        \vrule width.04\ht0}\hrule height.04\ht0}\mkern1.5mu}

\def\>{\rangle} 

\def\<{\langle} 
\def\Dsl{D \hskip-.6em \raise1pt\hbox{$ / $ } }

\def\sl#1{\rlap{\hbox{$\mskip 1 mu /$}}#1}
\def\leftrightarrowfill{$\mathsurround=0pt \mathord\leftarrow \mkern-6mu
       \cleaders\hbox{$\mkern-2mu \mathord- \mkern-2mu$}\hfill
       \mkern-6mu \mathord\rightarrow$}
\def\dvec#1{\vbox{\ialign{##\crcr
       \leftrightarrowfill\crcr\noalign{\kern-1pt\nointerlineskip}
       $\hfil\displaystyle{#1}\hfil$\crcr}}}          
\def\hook#1{{\vrule height#1pt width0.4pt depth0pt}}
\def\leftrighthookfill#1{$\mathsurround=0pt \mathord\hook#1
       \hrulefill\mathord\hook#1$}
\def\underhook#1{\vtop{\ialign{##\crcr                 
       $\hfil\displaystyle{#1}\hfil$\crcr
       \noalign{\kern-1pt\nointerlineskip\vskip2pt}
       \leftrighthookfill5\crcr}}}
\def\smallunderhook#1{\vtop{\ialign{##\crcr      
       $\hfil\scriptstyle{#1}\hfil$\crcr
       \noalign{\kern-1pt\nointerlineskip\vskip2pt}
       \leftrighthookfill3\crcr}}}

\def\sfrac#1#2{{\vphantom1\smash{\lower.5ex\hbox{\small$#1$}}\over
       \vphantom1\smash{\raise.4ex\hbox{\small$#2$}}}} 
\def\bfrac#1#2{{\vphantom1\smash{\lower.5ex\hbox{$#1$}}\over
       \vphantom1\smash{\raise.3ex\hbox{$#2$}}}}      
\def\afrac#1#2{{\vphantom1\smash{\lower.5ex\hbox{$#1$}}\over#2}}  
\def\on#1#2{{\buildrel{\mkern2.5mu#1\mkern-2.5mu}\over{#2}}}
\def\ddt#1{\on{\hbox{\LARGE .\kern-2pt.}}#1}             
\def\tdt#1{\on{\hbox{\LARGE .\kern-2pt.\kern-2pt.}}#1}   

\def\boxes#1{
       \newcount\num
       \num=1
       \newdimen\downsy
       \downsy=-1.5ex
       \mskip-2.8mu
       \bo
       \loop
       \ifnum\num<#1
       \llap{\raise\num\downsy\hbox{$\bo$}}
       \advance\num by1
       \repeat}
\def\boxup#1#2{\newcount\numup
       \numup=#1
       \advance\numup by-1
       \newdimen\upsy
       \upsy=.75ex
       \mskip2.8mu
       \raise\numup\upsy\hbox{$#2$}}

\newskip\humongous \humongous=0pt plus 1000pt minus 1000pt
\def\caja{\mathsurround=0pt}
\def\eqalign#1{\,\vcenter{\openup2\jot \caja
       \ialign{\strut \hfil$\displaystyle{##}$&$
       \displaystyle{{}##}$\hfil\crcr#1\crcr}}\,}
\newif\ifdtup

\def\to{\rightarrow}

\def\1ov4{{1\over 4}}

\def\tr{{\rm tr}}

\def\ddt{\dot{\t}}

\def\tr{\tilde{\r}}

\renewcommand{\d}{\delta}

\newcommand{\rmd}{{\rm d}}

\newcommand{\beq}{\begin{equation}}
\newcommand{\eeq}{\end{equation}}

\def\ba{\begin{eqnarray}}
\def\ea{\end{eqnarray}}

\def\tr{{\rm tr}}
\def\N{{\cal N}}
\def\J{{\cal J}}

\begin{document}


\null\vskip-24pt \hfill AEI-2000-067 \vskip-10pt\hfill
OHSTPY-HEP-T-00-023 \vskip-10pt 
\hfill
IFUM-FT-662 \vskip-10pt \hfill {\tt hep-th/0010137} \vskip0.2truecm
\begin{center}
\vskip 0.2truecm {\Large\bf
Perturbative and instanton corrections to the OPE of CPOs in
${\cal N}=4$ SYM$_4$
}\\
\vskip 0.5truecm
{\bf Gleb Arutyunov$^{*,**}$ \footnote{email:{\tt
agleb@aei-potsdam.mpg.de}}, Sergey Frolov$^{\ddagger ,**}$
\footnote{email:{\tt frolov@pacific.mps.ohio-state.edu}

$^{**}$On leave of absence from Steklov Mathematical Institute,
Gubkin str.8,
117966, Moscow, Russia }and Anastasios C. Petkou$^{\dagger}$
   \footnote{email:{\tt Anastasios.Petkou@mi.infn.it}
}
}\\
\vskip 0.4truecm
$^{*}$
{\it Max-Planck-Institut f\"ur Gravitationsphysik,
Albert-Einstein-Institut, \\
Am M\"uhlenberg 1, D-14476 Golm, Germany}\\
\vskip .2truecm $^{\ddagger}$ {\it Department of Physics,
The Ohio State University\\
Columbus, OH 43210-1106, USA}\\
\vskip .2truecm $^{\dagger}${\it Dipartimento di Fisica
dell'Universita di Milano,\\ Via Celoria 16,
20133 Milano, Italy}\\
\end{center}
\vskip 0.2truecm \noindent\centerline{\bf Abstract}
We study perturbative and instanton corrections to the Operator 
Product Expansion 
of the lowest weight Chiral Primary Operators 
of ${\cal N}=4$ SYM$_4$. 
We confirm the recently observed non-renormalization 
of various operators (notably
of the double-trace operator with dimension 4 in the ${\bf 20}$ irrep
of $SU(4)$),
that appear to be unprotected by unitarity restrictions.   
We demonstrate the splitting of the free-field theory 
stress tensor and R-symmetry
current in supermultiplets acquiring different 
anomalous dimensions in perturbation theory and 
argue that certain double-trace operators 
also undergo a perturbative splitting into operators 
dual to string and two-particle gravity states respectively.
The instanton contributions affect only those double-trace operators
that acquire finite anomalous dimensions at strong coupling. For the
leading operators of this kind, we show that
the ratio of their anomalous dimensions at strong coupling to 
the anomalous dimensions due to instantons is the same number.
\newpage

\section{Introduction}
${\cal N}=4$ supersymmetric Yang-Mills theory (SYM$_4$)
provides a concrete example of a
supersymmetric quantum field theory
where the idea of the AdS/CFT duality \cite{M,GKP,W} can be
successfully explored. According to the duality conjecture, SYM$_4$
with a gauge group $SU(N)$
at large $N$ and at strong 't Hooft coupling
$\lambda=g_{YM}^2N$ is dual to type IIB
supergravity on the $AdS_5\times S^5$ background.
Unifying the results obtained
in the context of the usual weak coupling expansion 
with the predictions of AdS/CFT duality allows us to
conceive basic dynamical features of the theory.

In the superconformal phase the non-trivial dynamics of the SYM$_4$
is encoded into correlation functions of gauge-invariant
composite operators, which may acquire perturbative as well as
non-perturbative (instanton) corrections.
An important class of local operators in SYM$_4$
is given by the Chiral Primary Operators (CPOs) of the form
$O_k^I=tr(\phi^{(i_1}\dots \phi^{i_k)})$,
where $\phi^{i}$ are the Yang-Mills scalars. Under supersymmetry
these operators generate short multiplets of the superconformal
algebra $SU(2,2|4)$ that are dual to multiplets of
type IIB supergravity compactified on  $AdS_5\times S^5$.
Unlike 2- and 3-point correlation functions of CPOs that are
subject to the known non-renormalization theorems
\cite{BaG,Lee,D'Hoker,Skiba,Gonzalez,PS,Penati1,Penati2,Howe},
4-point functions
receive in general perturbative and instanton corrections. As such, 
they contain important dynamical information for the
supersymmetry multiplets which appear in the Operator Product Expansion (OPE)
of two CPOs.

Recently, the 4-point function of the lowest weight CPOs $O_2^I$
has been computed in the supergravity approximation \cite{AF6,AF7},
and has been used
in \cite{AFP} to analyze
their  OPE at strong coupling.\footnote{Various aspects of 4-point
functions involving 
operators descendent to $O_2^I$ were discussed in \cite{LT1}-\cite{Hoffmann}.}
The structure of the OPE obtained in \cite{AFP} for the first few
low-dimensional operators was found to be in complete agreement
with the predictions of AdS/CFT correspondence.
Recall that
the transformation properties of local gauge--invariant operators
of SYM$_4$ with respect to the superconformal
algebra allow one to classify them into three categories:

\noindent {\it i}) ``Single-trace'' chiral operators which belong
to short representations and have  conformal dimensions protected
from quantum corrections.

\noindent {\it ii}) Operators which are
obtained as ``normal-ordered'' products of the chiral operators.
They may belong
either to short or long representations,
the former have protected conformal dimensions, while the dimensions
of the latter are restricted from above.

\noindent {\it iii}) Operators
which belong to long representations and whose conformal dimensions
grow without bound in the strong coupling limit.

According to the  AdS/CFT duality,  the operators in {\it i})
are dual to the type IIB supergravity fields while operators in
{\it ii}) are dual to multi-particle
supergravity states. For the operators in 
{\it iii}) the duality predicts the growth
of their conformal dimensions as $\lambda^{1/4}$ when
$\lambda\rightarrow \infty$. The latter operators are interpreted 
as being dual to string states (single- or multi-particle),
which decouple in the strong coupling limit.

Comparison of the OPE of the two lowest weight CPOs in free-field
theory and at strong 
coupling \cite{AFP} 
has enabled us to make the
following predictions for the structure of the OPE at finite
$\lambda$ and $N$:

\noindent $i$) The $R$-symmetry current and the
stress tensor of the free-field theory, which involve   
only the six SYM scalars $\phi^i$,  undergo splitting into 2 and 3
operators respectively belonging to different supermultiplets.
Only one operator in each splitting is dual to a supergravity field
and has protected conformal dimension, while all others decouple at strong
coupling as their anomalous dimensions grow without bound.

\noindent $ii$) The only double-trace operator with free-field
conformal dimension 4 that acquires an anomalous dimension at strong
coupling is
$O_1 = :O^IO^I:+\cdots$ transforming in the trivial representation
of the $R$-symmetry group. We argue that the free-field theory 
operator $O_1^{fr}$
also undergoes splitting into a sum of an operator dual to a gravity
state and operators dual to string modes. The same kind of
splitting also occurs for the scalar operator in the ${\bf 84}$ irrep.
The double-trace scalar operators in the
${\bf 20}$ and ${\bf 105}$ irreps do not split. 
The operator in the ${\bf 20}$ irrep saturates  
the unitarity bound A') in the classification of \cite{AFer2}
and is not protected from acquiring anomalous dimension.
However, our analysis shows that this operator 
retains its canonical dimension
and hence its non-renormalization is a genuine dynamical effect.

\noindent $iii$) The double-trace operator with free-field dimension 5
in the ${\bf 15}$
irrep acquires anomalous dimension, while the one in the ${\bf 175}$
has protected dimension. They both split 
at finite $\lambda$ and $N$.

\noindent $iv$) There are several towers of traceless symmetric
tensor operators in the
${\bf 105}$, ${\bf 84}$ and ${\bf 175}$  irreps, whose anomalous
dimensions vanish.

Here we confirm the above  predictions by analyzing the
4-point function of the CPOs $O_2^I$ computed at 2-loops in
perturbation theory \cite{GPS,EHSSW0,EHSSW,BKRS1} (three-loop
results were obtained in  
\cite{ESS,BKRS2}). We also study the 
instanton contribution to the 4-point function. In \cite{BGKR}
the correlation functions of the four
${\cal N}=2$ singlet scalar fields and of the sixteen dilatinos
were computed in SYM$_4$ with gauge group $SU(2)$ in
the sector with instanton number $k=1$.
In \cite{DKMV} these results were further generalized to
the group $SU(N)$ and in \cite{DHKMV1,DHKMV2} to arbitrary $k$
in the large $N$ limit. With the above results at hand, we then use
the recently obtained non-renormalization theorem of \cite{EPSS} to
restore the {\it complete}  4-point function of the CPOs $O_2^I$ and perform its
OPE analysis. Our results are in agreement with the earlier
considerations of \cite{IS} 
 and  
show the absence of instanton contributions to the anomalous dimensions 
of single-trace operators in the Konishi multiplet \cite{BKRS1}.
Pictorially, we observe that the instanton contribution 
is ``seen'' only by those operators whose 
anomalous dimensions are non-zero and finite at strong coupling. In
particular, the double-trace operator in $\bf 20$ does not receive  
instanton corrections. Such
a picture points to an interesting relation between the mysterious
``multi-particle'' supergravity states and the D-particle modes.

The plan of the paper is as follows. 
In Section 2 we recall the OPE structure of two 
CPOs at weak and strong coupling. In Section 3 we analyze the
two-loop 4-point function of the lowest weight CPOs. 
We compute the anomalous dimensions of
single- and double-trace operators and demonstrate the splitting of
the free-field operators into distinct supermultiplets acquiring
different anomalous dimensions. In Section 4 we study the instanton
contribution to the 4-point function of the CPOs and show that
instantons do not contribute to the anomalous 
dimensions of neither operators dual to string-modes nor operators
with protected dimension. The dynamically protected operator in the
${\bf 20}$ does not receive instanton contributions, which indicates
that only operators receiving finite-anomalous dimensions at strong
coupling ``see'' instantons. 
In the conclusion we discuss the results
obtained.

\section{OPE algebra of CPOs at weak and strong coupling}
In this Section we review the structure of the OPE algebra of the lowest weight
CPOs at both the weak and the strong coupling regimes. We follow the notation
of \cite{AFP} and also use 
\beq \label{tlambda} \tilde{\lambda} = \frac{\lambda}{(2\pi)^2}
=\frac{g_{YM}^2 N}{(2\pi)^2}\, .\eeq

The normalized lowest weight CPOs in SYM$_4$ are operators of the form
$$
O^I(x)=\frac{1}{2^{1/2}\tilde{\lambda}}C_{ij}^I \tr(:\phi^i\phi^j:),
$$
where the symmetric traceless
tensors $C_{ij}^I$, $i,j=1,2,..,6$ form an orthonormal basis of the
${\bf 20}$ of $SO(6)$.
As was shown in \cite{AFP}
the leading terms in the OPE of two $O^I$s in free field theory
take the form
\ba
O^{I_1}(x_1)O^{I_2}(x_2)&=&\frac{\d^{I_1I_2}}{x_{12}^4}
+\frac{2^{3/2}}{N}\frac{C^{I_1I_2I}}{x_{12}^2}[\,O^I\,]
+\frac{2}{3^{1/2} N}
\frac{\d^{I_1I_2}}{x_{12}^2}[\,{\cal K}\,] \nonumber \\
&&\hspace{-1cm}+\, \frac{2^{3/2}}{\tilde{\lambda} N}\frac{x_{12}^{\mu}}{x_{12}^2}
C^{I_1I_2}_{\J_{15}}[\,J^{\J_{15}}_{\mu}\,]
-\frac{\d^{I_1I_2}}{6\tilde{\lambda} N }
\frac{x_{12}^{\mu}x_{12}^{\nu}}{x_{12}^2}[\,T_{\mu\nu}^{free}\,]
+\frac{1}{\tilde{\lambda} N}
\frac{x_{12}^{\mu}x_{12}^{\nu}}{x_{12}^2}C^{I_1I_2I}
[\,T_{\mu\nu}^{I}\,]\nonumber \\
&&\hspace{-1cm}+\,\d^{I_1I_2}[\,O_{1}\,]+
C^{I_1I_2}_{{\cal J}_{20}}[\,O^{{\cal J}_{20}}\,]
+C^{I_1I_2}_{{\cal J}_{105}}[\,O^{{\cal J}_{105}}\,]+
C^{I_1I_2}_{{\cal J}_{84}}[\,O^{{\cal J}_{84}}\,]\nonumber \\
&&\hspace{-1cm}+\, C^{I_1I_2}_{{\cal J}_{15}}x^\mu_{12}[\,O_\mu^{{\cal
      J}_{15}}\,]
+C^{I_1I_2}_{{\cal J}_{175}}x^\mu_{12}[\,O_\mu^{{\cal J}_{175}}\,]+\ldots .
\label{freeope}
\ea
Here $T_{\mu\nu}^{free}$ and $J^{\J_{15}}_{\mu}$ are respectively the stress tensor and
the normalized $R$-symmetry current of the
free field theory (including only six scalar
fields), ${\cal K}$ is the normalized Konishi scalar, $T_{\mu\nu}^I$
is a traceless second rank tensor in ${\bf 20}$ and $O^{\cal J}$ denote
generically double-trace operators in the corresponding representation
${\cal J}$ 
of the R-symmetry. In addition to the above fields the OPE contains
infinite towers of both single-trace as well as double-trace operators.

The strong coupling OPE compatible with the 4-point function
of \cite{AF6} is different from (\ref{freeope}) and reads
\ba
O^{I_1}(x_1)O^{I_2}(x_2)&=&\frac{\d^{I_1I_2}}{x_{12}^4}
+\frac{2^{3/2}}{N}\frac{C^{I_1I_2I}}{x_{12}^2}[\,O^I\,]
+\frac{2^{3/2}}{3\tilde{\lambda} N}\frac{x_{12}^{\mu}}{x_{12}^2}
C^{I_1I_2}_{\J_{15}}[\,R^{\J_{15}}_{\mu}\,]\nonumber \\
&-&\frac{1}{30\tilde{\lambda} N }\d^{I_1I_2}
\frac{x_{12}^{\mu}x_{12}^{\nu}}{x_{12}^2}[\,T_{\mu\nu}\,]
+\d^{I_1I_2}x^{\D_{1}^{(s)}}_{12}[\,O_{1}\,]\nonumber \\
&+&
C^{I_1I_2}_{{\cal J}_{20}}x^{\D_{20}^{(s)}}_{12}
[\,O^{{\cal J}_{20}}\,]
+C^{I_1I_2}_{{\cal J}_{105}}x^{\D_{105}^{(s)}}_{12}
[\,O^{{\cal J}_{105}}\,]
+
C^{I_1I_2}_{{\cal J}_{84}}x^{\D_{84}^{(s)}}_{12}
[\,O^{{\cal J}_{84}}\,]\nonumber \\
&+&C^{I_1I_2}_{{\cal J}_{15}}x^{\D_{15}^{(s)}}_{12}x^\mu_{12}
[\,O_\mu^{{\cal J}_{15}}\,]
+C^{I_1I_2}_{{\cal J}_{175}}x^{\D_{175}^{(s)}}_{12}x^\mu_{12}
[\,O_\mu^{{\cal J}_{175}}\,]+\ldots .
\label{strongope}
\ea
Here $R^{\J_{15}}_{\mu}$ is the R-symmetry current and $T_{\mu\nu}$ is
the stress tensor of the {\it full} $\N =4$ SYM$_4$ and $\D_{\cal J}^{(s)}$
is the anomalous dimension of the corresponding double-trace operator
at strong coupling. The conformal blocks appearing in
(\ref{strongope}) encode all the strong coupling information for the anomalous dimensions
and the couplings of the corresponding operators. 
In the place of an infinite number of
single-trace operators in (\ref{freeope}), (\ref{strongope})
contains instead only three
single-trace operators
giving rise to the most singular terms. Note that the
coefficients in front of the $R$-symmetry current and the stress
tensor in (\ref{strongope}) are different from the ones in (\ref{freeope}).
The reason is
that the free-field operators $J_\mu^{\J_{15}}$ and $T_{\mu\nu}^{free}$
constructed only from scalars are split into operators
belonging to different supersymmetry multiplets. Multiplets that are dual
to string modes decouple in the strong coupling limit, while operators
from the stress tensor multiplet are non-renormalized
and show up at strong coupling.

The leading double-trace operators 
receive anomalous dimensions whose value at strong coupling
was found to be
\bea
\D_{O_{\bf 1}}^{(s)}=-\frac{16}{N^2},~~~\D_{O_{\bf 15}}^{(s)}=-\frac{16}{N^2},
\eea
while all the other operators shown in (\ref{strongope})
have vanishing anomalous dimensions.
The double trace operators in {\bf 84}, {\bf 105}
are in short multiplets and they are protected.
The double-trace operator in {\bf 20} is not protected by unitarity
and is allowed to acquire an anomalous dimension. 
Nevertheless, it was found to have vanishing anomalous 
dimension at strong coupling.

Comparison of the free-field  and strong coupling OPEs (\ref{freeope})
and (\ref{strongope}) 
enabled us to make the predictions for the OPE structure at finite $N$
and $\lambda$ discussed in the Introduction. In the next
Section we verify that these predictions are in agreement
with the 2-loop 4-point function of CPOs.
To this end we study the asymptotic behavior of the 4-point
function in the direct channel $x_{12}^2,x_{34}^2 \to 0$,
which in terms of the ``biharmonic ratios''
$
u=\frac{x_{12}^2x_{34}^2}{x_{13}^2x_{24}^2}
\,,v=\frac{x_{12}^2x_{34}^2}{x_{14}^2x_{23}^2}
$
and the variable $Y=1-\frac{v}{u}$, amounts to taking the  short-distance limit
$u,\,v,\,Y\,\rightarrow 0$.
Our analysis closely follows \cite{AFP} and is based on
the knowledge of the conformal partial wave
amplitudes of quasi-primary operators.
In particular, consider the contributions to the OPE of two CPOs
coming from  
a scalar, vector and second rank
symmetric traceless tensor. Schematically this is given by
\ba O^{I_1}(x_1)O^{I_2}(x_2)&=& C^{I_1I_2}_{{\cal
J}}\biggl( \frac{C_{OOS}}{C_{S}}
\frac{1}{x^{4-\D_S}_{12}}[\,S^{{\cal J}}\,] -\frac{C_{OOT}}{C_{T}}
\frac{x_{12}^\mu x_{12}^\nu}{x^{6-\D_T}_{12}}
[\,T^{{\cal J}}_{\mu\nu}\,]\nonumber \\
&~&~~~~~~+ \frac{C_{OOV}}{C_{V}} \frac{x_{12}^\mu
}{x^{5-\D_V}_{12}} [\,V^{{\cal J}}_{\mu}\,]+\ldots \biggr) .
\label{opestv}
\ea
Here $\J$ denotes an index of an irrep
of the $R$-symmetry group $SO(6)$,
$C^{I_1I_2}_{{\cal J}}$ are the Clebsch-Gordan coefficients  and
$\D_S,\,\D_T,\,\D_V$ are the conformal dimensions of the scalar,
tensor and vector operators  respectively.
For any 
operator in the OPE, $C_{\cal O}$ and $C_{OO{\cal O}}$ denote the
normalization constant in the 2-point function $\langle{\cal
O}(x_1){\cal O}(x_2)\rangle$ and the coupling constant in the
three-point function $\langle O^I(x_1)O^J(x_2) {\cal
O}(x_3)\rangle$, respectively.
Then, the  short-distance expansion of the conformal partial amplitudes (CPWA) of
the scalar S, tensor T and vector V operators can be written as ({\it c.f.}
\cite{AFP})
\ba
&&\langle
  O^{I_1}(x_1)O^{I_2}(x_2)O^{I_3}(x_3)
  O^{I_4}(x_4)\rangle = \frac{C^{I_1I_2}_{{\cal J}}C^{I_3I_4}_{{\cal J}}}
{x_{12}^4x_{34}^4}
\nonumber \\
&&\hspace{1.5cm}\times \Biggl[
\frac{C_{OOS}^2}{C_S}v^{\frac{\Delta_S}{2}} \Biggl(
1+\frac{\Delta_S}{4}Y +\frac{\Delta_S^3}{16(\Delta_S
-1)(\Delta_S+1)}v\left( 1 +\frac{\Delta_S+2}{4}Y\right)
+\cdots\Biggl)
\nonumber \\
&&
\hspace{3cm}+\frac{C_{OOT}^2}{C_T}v^{\frac{\Delta_T}{2}-1}\Biggl(
\frac{1}{4}Y^2
-\frac{1}{4}v-\frac{\Delta_T}{16}vY\cdots\Biggl)
\nonumber \\
&& \hspace{3.5cm}+\frac{C_{OOV}^2}{C_V}v^{\frac{\Delta_V
-1}{2}}\Biggl( \frac{1}{2}Y +\cdots\Biggl)\Biggl] ,
\label{OPEa}
\ea
where we assumed that $\Delta_T = 4 + \Delta_T^{(1)}$ and kept only terms
linear in $\Delta_T^{(1)}$.
The formulas for the leading contributions of a rank-2
traceless symmetric tensor and a vector can be generalized to the
case of a rank-$l$ traceless symmetric tensor of dimension $\D_l$
and one gets a leading term of the form
$v^{\frac{\Delta_l -l}{2}}\,Y^l$.

If we decompose the conformal dimension of an operator into a
``canonical'' part (equal to its free-field conformal dimension) and
an ``anomalous'' part, taken to be a small parameter (see \cite{AFP}),
then (\ref{OPEa})
shows that the ``anomalous dimensions'' are related to terms of the
form $v^{\frac{\D_S^{(0)}}{2}}\log v$ for scalar operators,
$v^{\frac{\D_V^{(0)}-1}{2}}\,Y\,\log v$ for vector operators and
$v^{\frac{\D_T^{(0)}-2}{2}}\,Y^2\,\log v$ for rank-2 tensor
operators. Formula (\ref{OPEa}) is the basic tool in our analysis
of the 2-loop 4-point function in Section 3 and the instanton
contribution in Section 4.

\section{OPE analysis of the 2-loop 4-point function}
The 2-loop 4-point functions of the CPOs $O_2$ were computed in \cite{GPS,EHSSW0,EHSSW}
and the results obtained there can be represented in terms of a basic function
$\Phi^{(1)}(v,u)$ that can be written in the form of a Mellin-Barnes
integral as
\beq
\label{defPhi}
\Phi^{(1)}(x,y) = \frac{1}{(2\pi{\rm i})^2} \int_{\cal C} \rmd s\,\rmd
t \,\Gamma^2(-s)\Gamma^2(-t)\Gamma^2(1+s+t)\,x^s\,y^t \, ,
\eeq
where the contour(s) ${\cal C}$ run parallel to the imaginary
axis. Performing the integrations, we may cast it in a form suitable for
studying the OPE as
\ba
\label{resPhi}
\Phi^{(1)}(v,Y)
&=& \sum_{n,m=0}^{\infty} \frac{v^n Y^m}{(n!)^2 m!}
\frac{\Gamma^2(1+n) \Gamma^2(1+n+m)}{\Gamma(2+2n+m)}\nonumber \\
&&\hspace{1cm}\times\left[-\log
  v+2\psi(2+2n+m) -2\psi(1+n+m)\right] \, .
\ea
Then, the 4-point function of the CPOs $O^I$ reads
\ba
\label{2loop4pt}
\langle O^{I_1}(x_1) O^{I_2}(x_2)O^{I_3}(x_3)O^{I_4}(x_4)\rangle &=&
\d^{I_1I_2}\d^{I_3I_4}a_1
+\d^{I_1I_3}\d^{I_2I_4}a_2
+\d^{I_1I_4}\d^{I_2I_3}a_3\nonumber \\
&&\hspace{-1cm}
+C^{I_1I_2I_3I_4}b_2+C^{I_1I_3I_2I_4}b_1+C^{I_1I_3I_4I_2}b_3\, ,
\ea
where up to 2-loops the various coefficients are given by
\bea
\label{1la1b3}
\begin{array}{ll}
a_1  =  \frac{1}{x_{12}^4
  x_{34}^4}\left[1-\frac{2\tilde{\lambda}}{N^2}v
  \Phi^{(1)}(v,Y) \right] \, ,
&
b_1  =  \frac{4}{N^2}\frac{1}{x_{12}^4 x_{34}^4}
\left[vu-\frac{\tilde{\lambda}}{2} 
v(vu-v-u)
  \Phi^{(1)}(v,Y) \right]\, ,  \\
a_2  =  \frac{1}{x_{12}^4 x_{34}^4} \left[u^2-\frac{2\tilde{\lambda}}{N^2}vu
  \Phi^{(1)}(v,Y) \right], 
& b_2 =  \frac{4}{N^2}\frac{1}{x_{12}^4
  x_{34}^4}\left[v+\frac{\tilde{\lambda}}{2}v(v+Y)
  \Phi^{(1)}(v,Y) \right ]\,, \nonumber \\
a_3  =  \frac{1}{x_{12}^4 x_{34}^4} \left[v^2-\frac{2\tilde{\lambda}}{N^2}v^2
  \Phi^{(1)}(v,Y) \right]\, ,
& b_3  =  \frac{4}{N^2}\frac{1}{x_{12}^4
  x_{34}^4}\left[u+\frac{\tilde{\lambda}}{2}\frac{v-Y}{1-Y}
  v\Phi^{(1)}(v,Y) \right] \, .\nonumber 
\end{array}
\ea 
Using the above result we may now study the OPE at 2-loops. We start
with the projection into the singlet which includes important fields
such as the stress tensor, the Konishi scalar and the
double-trace operator $O_1$ with canonical dimension 4.

\vskip 0.4cm
{\it 3.1 Projection in the singlet }
\vskip 0.4cm

Using the properly normalized projector in the singlet \cite{AFP}, we obtain
for the first few terms in the short-distance expansion
\ba
\nonumber &&\langle
O^{I_1}(x_1)O^{I_2}(x_2)O^{I_3}(x_3)O^{I_4}(x_4) \rangle |_{{\bf 1
}} = \frac{\d^{I_1I_2}\d^{I_3I_4}}{x_{12}^4x_{34}^4}\Biggl[
1+\frac{4}{3N^2}v\left(1+\frac{3\tilde{\lambda}}{2}\log
  v-3\tilde{\lambda}\right) \\
&&\hskip 1cm +\frac{2}{3N^2}vY\left(1+\frac{3\tilde{\lambda}}{2}\log v
  -\frac{3\tilde{\lambda}}{2} \right)
 + \frac{2}{3N^2}vY^2\left( 1+\frac{3\tilde{\lambda}}{2}\log
  v-\frac{5\tilde{\lambda}}{3} \right) \label{p0exp} \\
&&+\frac{1}{10}v^2\left(
    1+\frac{2}{3N^2} -\frac{2\tilde{\lambda}}{N^2}\log
    v+\frac{230\tilde{\lambda}}{45N^2}\right)
 +\frac{1}{10}v^2 Y\left(1+\frac{2}{3N^2}
  -\frac{2\tilde{\lambda}}{N^2}\log
  v+\frac{37\tilde{\lambda}}{9N^2}\right) \nonumber \Biggr] \, .
\ea
The expansion (\ref{p0exp}) should be matched with the contributions
coming from the first
few low-dimensional operators in the singlet projection of the OPE (\ref{freeope}). In the
free-field theory limit the first fields which appear in the above OPE 
are the Konishi scalar ${\cal K}$ with free-field dimension 2,
the stress tensor of 6 free scalar fields $T_{\m\n}^{free}$ and a double-trace
operator $O_1^{free}$ with free-field dimension 4. It is natural to assume
that these are exactly the first few operators which appear also in the 2-loop OPE,
allowing only for possible small corrections in their free-field dimensions and
normalization constants in order to account for the logarithmic terms in (\ref{p0exp}).
Although such an assumption seems to work for the Konishi scalar, it does not
quite fit the 2-loop result (\ref{p0exp}) as there is a discrepancy in
the coefficients in front of the stress tensor in free field theory
and at 2-loops. 

In order to properly interpret (\ref{p0exp}) one should realize
that the stress tensor $T_{\m\n}$ expected to appear in it is
different from $T_{\m\n}^{free}$, since it 
receives contributions not only from the six scalars but also from the
four Weyl fermions and the vector field of ${\cal N}=4$ SYM. It has been
argued in \cite{An}, following \cite{OP},  that the general stress
tensor of an interacting CFT involving scalars, fermions and vectors involves
three, linearly independent and mutually orthogonal
structures.\footnote{This is easily seen in 
free-field theory, where the stress tensors for free scalars, fermions
and vectors provide three linearly independent and orthogonal to each
other structures \cite{OP}.} For the specific case of ${\cal N}=4$
SYM, simple manipulations allow one to write the free-field
 stress tensor (i.e. the stress tensors of six massless free scalars)
$T_{\m\n}^{free}(x)$ as follows  
\ba \label{tsplit}
T_{\m\n}^{free}(x) &=& \frac{1}{5}T_{\m\n}(x)
+\frac{10}{35}{\cal K}_{\m\n}(x) +\frac{18}{35}\Xi_{\m\n}(x) \, ,\ea
where the three structures depicted in (\ref{tsplit}) are mutually
orthogonal and linearly independent. The idea of \cite{An} is that the
orthogonality and linear independence property is preserved by
perturbation theory, i.e. the structures in (\ref{tsplit}) do not mix
under renormalization. 
The (symmetric and traceless) tensor ${\cal K}_{\m\n}$
belongs to the Konishi supermultiplet while the (symmetric and
traceless) $\Xi_{\m\n}$ is the lowest component of a new
supermultiplet. The full stress tensor $T_{\m\n}$ is expected to
remain conserved, therefore, it retains its canonical dimension at any
order in perturbation theory. However, ${\cal K}_{\m\n}$ and
$\Xi_{\m\n}$ can, and do, acquire 
anomalous dimensions.

We preface the more detailed analysis of
the OPE with some necessary comments.
The fact that $T_{\mu\nu}$ is canonically
normalized allows one to find the free-field value of 
the normalization constants
of the 2-point functions of the three operators in (\ref{tsplit}) as
\bea
C_T=5C_t,~~~C_{{\cal K}_1}=\frac{7}{2}C_t,~~~ C_{\Xi} =\frac{35}{18}C_t,
\label{nc}
\eea
where $C_t=32\tilde{\lambda}^2$ is a normalization constant for
$T_{\m\n}^{free}$. The value of the coupling
$C_{OOT}=\frac{16\tilde{\lambda}}{3N}$ is
fixed by the conformal Ward identity. The 
free-field theory OPE (\ref{freeope}) together
with (\ref{tsplit}) and (\ref{nc}) gives the free-field value of the normalization
constants
\bea
C_{OO{\cal K}_{\bf 1}}=C_{OO\Xi}=\frac{16\tilde{\lambda}}{3N}\, .
\nonumber
\eea
Recall that the Konishi field is canonically normalized, i.e. $C_{\cal K}=1$ and
the free-field result for $C_{OO{\cal K}}$ is $C_{OO{\cal
K}}=\frac{2}{3^{1/2}N^2}$.

In the sequel we assume that for any operator $\cal O$ in the OPE  
the ratio $\frac{C_{OO{\cal O}}}{C_{{\cal O}}}$
is kept equal to its free-field value. The correction to a coupling dependent
normalization constant $C_{OO{\cal O}}(\tilde{\lambda})$ is introduced 
in the following way 
\bea
C_{OO{\cal O}}(\tilde{\lambda})=C_{OO{\cal O}}(1+C_{OO{\cal O}}^{(1)})\, ,
\eea 
where $C_{OO{\cal O}}$ stands for the free-field value.

Now taking into account the splitting (\ref{tsplit}), using
(\ref{OPEa}) for the contributions of scalars and symmetric traceless
tensors to the OPE and expanding the anomalous dimensions and the normalization
constants, we find for the leading terms of the short-distance expansion of the
singlet projection  
\ba \label{p0ope} P_{singlet}^{OPE}
=\frac{\d^{I_1I_2}\d^{I_3I_4}}{x_{12}^4 x_{34}^4}\left[1+A_{10}v + A_{11}vY
  +A_{12}vY^2 
\right]\, ,
\ea where the coefficients $A$ are given by
\ba
\label{A10} A_{10}&=& \frac{C_{OO{\cal K}}^2}{C_{\cal K}}\left[1+\frac{\eta_{\cal
K}}{2}\log
  v+C_{OO{\cal K}}^{(1)}\right] \, ,\\
\label{A11} A_{11}&=& \sfrac12\frac{C_{OO{\cal K}}^2}{C_{\cal
K}}\left[1+\frac{\eta_{\cal K}}{2}\log
  v+\frac{\eta_{\cal K}}{2}+C_{OO{\cal K}}^{(1)}\right] \, ,\\
\label{A12} A_{12}&=& \sfrac{1}{3}\frac{C_{OO{\cal K}}^2}{C_{\cal
K}}\left[1+\frac{\eta_{\cal K}}{2}\log
  v+C_{OO{\cal K}}^{(1)} +\sfrac{2}{3}\eta_{\cal K}\right]
+\sfrac{1}{4}\frac{C_{OOT}^2}{C_T} \\
&+&\sfrac{1}{4}\frac{C_{OO{\cal K}_{\bf 1}}^2}{C_{{\cal K}_{\bf 1}}}
\left[1+\sfrac{1}{2}\eta_{{\cal K}_{\bf 1}}\log
  v+C_{OO{\cal K}_{\bf 1}}^{(1)}\right]
 +\sfrac{1}{4}\frac{C_{OO\Xi}^2}{C_{\Xi}}\left[1+\sfrac{1}{2}\eta_{\Xi}\log
  v+C_{OO\Xi}^{(1)}\right] \nonumber \, .
\ea
Here the parameters  $\eta_{\cal K}$, 
$\eta_{{\cal K}_{}\bf 1}$ and $\eta_{\Xi}$ correspond to 
the small corrections to the canonical
dimensions of the operators $\cal K$, ${\cal K}_{\m\n}$ and
$\Xi_{\m\n}$, respectively, while $C_{OO{\cal K}}^{(1)}$, 
$C_{OO{\cal K}_{\bf 1}}^{(1)}$ and $C_{OO\Xi}^{(1)}$
denote the small corrections to the corresponding free-field normalization constants. 
According to the discussion above the free-field values of the ratios of the 3-
and 2-point normalization constants occurring in (\ref{A10})-(\ref{A12}) are
given by
\bea \frac{C_{OO{\cal K}}^2}{C_{\cal K}}=\frac{4}{3N^2}\, , ~~~~
\frac{C_{OO{\cal K}_{\bf 1}}^2}{C_{{\cal K}_{\bf 1}}}=\frac{16}{63N^2}\, ,~~~~
\frac{C_{OO\Xi}^2}{C_{\Xi}}=\frac{16}{35N^2}\, .
 \eea
Requiring consistency of the terms carrying equal powers of $v$ and
  $Y$ in (\ref{p0exp}) and 
(\ref{p0ope}) 
 we then obtain the anomalous dimensions and corrections to
the coupling constants of the operators discussed above.

Consistency of the terms proportional to $v$ in (\ref{p0exp}) and
(\ref{p0ope}) gives
\beq
\label{ankonishi}
\eta_{\cal K} =
3\tilde{\lambda} \,,~~~~~~~C_{OO{\cal K}}^{(1)} =-3\tilde{\lambda} \, .
\eeq
The value of $\eta_{\cal K}$ coincides with the well-known
value for the 2-loop anomalous dimension for the Konishi scalar
\cite{An}.

Using the result (\ref{ankonishi}), we immediately see that the terms
proportional to $vY$ in (\ref{p0exp}) and (\ref{p0ope}) are consistent.

Consistency of the terms proportional to $vY^2$ in (\ref{p0exp}) and
(\ref{p0ope}) gives
\ba
\label{vYY1}
\frac{3}{2}\tilde{\lambda} &=& \frac{1}{7}\eta_{{\cal
K}_1}
+\frac{9}{35}\eta_{\Xi}\, , \\
\label{vYY2} -\frac{3}{2}\tilde{\lambda}&=&
\frac{1}{7}C_{OO{\cal K}_{1}}^{(1)}
+\frac{9}{35}C_{OO\Xi}^{(1)}\, .
\ea
This shows that the consistency of
the short-distance expansion with the OPE
is not sufficient to determine the
individual anomalous dimensions and corrections to the couplings of the split
fields ${\cal K}_{\m\n}$ and $\Xi_{\m\n}$. 
However, here comes the input of supersymmetry
which rectifies the situation. Namely, requiring that ${\cal K}_{\m\n}$
is in the same supermultiplet as the Konishi scalar $\cal K$ we fix its anomalous
dimension to be the same as $\cal K$
\beq
\eta_{{\cal K}_{\bf 1}}=3\tilde{\lambda} \, .
\eeq
Then, we easily find from (\ref{vYY1}) that
\beq
\eta_{\Xi} =
\frac{25}{6}\tilde{\lambda} \, .
\eeq
in complete agreement with \cite{An}. 

The terms in (\ref{p0exp}) proportional to $v^2$ and $v^2Y$ encode the 
information about scalar operators of free-field dimension 4. Recall that in free-field 
theory the corresponding terms match with the contribution of a unique operator 
$$
O_1^{free}=\frac{1}{40\tilde{\lambda}^2}\left(:\tr(\phi^i\phi^j)\tr(\phi^i\phi^j):-
\frac{1}{6}:\tr(\phi^i\phi^i)\tr(\phi^j\phi^j):\right)
$$
with the 2-point function
\bea
\label{Ofree}
\langle
O^{free}_1(x_1)O^{free}_1(x_2)\rangle =
\frac{1}{10}\left(1+\frac{2}{3N^2}\right) \frac{1}{x_{12}^8} \, ,
\eea
while at strong coupling the singlet channel was shown \cite{AFP}
to contain a scalar operator $O_1$ of approximate dimension 4 
with the following 2-point function: 
\bea
\label{Ostrong}
\langle
O_1(x_1)O_1(x_2)\rangle =
\frac{1}{10}\left(1+\frac{38}{15N^2}\right) \frac{1}{x_{12}^{8+2\Delta_1^{(s)}}} \, .
\eea
Although the difference of 2-point functions of
$O_1^{free}$ and $O_1$ might be explained
by the fact that they  are computed in
different regimes and the operator is not protected, it is more natural 
to assume that $O_1^{free}$ splits in perturbation theory into a sum of
operators such that  only one of them is dual to a gravity state.  
Indeed, in free-field theory one finds a number of linearly independent 
operators of dimension 4, e.g.,  $:\tr(\phi^i\phi^j)\tr(\phi^i\phi^j):$, 
$:\tr(\phi^i\phi^i)\tr(\phi^j\phi^j):$ and 
$:\tr(\phi^i\phi^j\phi^i\phi^j):$ that may mix under renormalization. 
To find the individual anomalous dimensions at two loops one should  
diagonalize their mixing matrix. We then expect to find a unique operator $O_1$ 
(dual to a ``two-particle'' gravity state), whose anomalous 
dimension behaves as $\frac{\tilde{\lambda}}{N^2}$, while the other 
operators (dual to string modes) should have the 
anomalous dimensions of the Konishi type
$\sim\tilde{\lambda}$. Such a splitting, similar in spirit with
the above discussed splitting of the stress tensor, seems to be
necessary in order to account for the fact that at strong coupling
we find only one operator with approximate dimension 4 while at any
order in perturbation theory we expect an operator mixing. However,
the knowledge of the    
correlation functions of CPOs alone is not sufficient in order to
establish the mixing 
matrix and additional information  is needed, e.g. the knowledge of
correlation functions of four Konishi scalars or other operators.

\vskip 0.4cm
{\it 3.2 Projection in {\bf 20}}
\vskip 0.4cm

Projecting the 4-point function in the  ${\bf 20}$ irrep we obtain
for the leading in $v,Y$ terms the following answer
\bea
\label{p20}
&&\langle
O^{I_1}(x_1)O^{I_2}(x_2)O^{I_3}(x_3)O^{I_4}(x_4) \rangle
|_{{\bf20}} = \frac{C_{{\cal J}_{20}}^{I_1I_2}C_{{\cal
J}_{20}}^{I_3I_4}}{x_{12}^4x_{34}^4} \Biggl[
\frac{40}{3N^2}v\left(1+\frac{1}{2}Y+\frac{1}{2}Y^2 \right)+ \\
\nonumber && +v^2\left(2+\frac{2}{3N^2}\right)(1+Y)
-\frac{\tilde{\lambda}}{N^2}\frac{20}{3}v\left(Y^2-v-\frac{3}{4}vY\right)
+\frac{\tilde{\lambda}}{N^2}\frac{10}{3}v\left(Y^2-v-vY\right)\log{v}
 \Bigg]\, .
 \eea
According to our discussion of the free-field theory OPE in Section 2, the first
three low-dimension operators contributing to ${\bf 20 }$ are the CPOs
themselves, the
double-trace operator $O_{\bf 20}$, and a symmetric second rank tensor ${\cal
K}_{\m\n}^{I}\equiv {\cal K}_{\bf 20}$ of approximate dimension 4. On the other
hand at strong coupling and in the large $N$ limit we found that only
the CPOs and the 
$O_{\bf 20}$ survive and keep their free-field dimension. While
the non-renormalization property of CPO is well-known, the non-renormalization of $O_{\bf
20}$ is a new phenomenon that cannot be explained on the basis  of unitarity. A
natural suggestion made in \cite{AFP} is that $O_{\bf 20}$ is non-renormalized in
perturbation theory at finite $N$. As far as ${\cal K}_{\bf 20}$ is concerned, being dual to a
string mode  it receives perturbatively large anomalous dimension and
decouples from the spectrum at strong coupling.\footnote{In principle one could
expect a splitting of ${\cal K}_{\bf 20}$ into a sum of operators
dual to string modes. However, our analysis will show that this does not happen.}
Let us see how this picture is
compatible with two-loop result (\ref{p20}).

The last two terms in (\ref{p20}) are proportional to $\tilde{\lambda}$ 
and we interpret
them as loop contribution to the coupling 
$C_{OO{\cal K}_{\bf 20}}$ and 
to the anomalous dimension of ${\cal K}_{\bf 20}$ respectively.
Indeed, if we denote the anomalous dimension of ${\cal K}_{\bf 20}$ as
$\Delta_{{\cal K}_{\bf 20}}^{(1)}$ then the $\log{v}$-dependent term in
(\ref{p20}) occurs due to the contribution of the conformal block of the second
rank tensor with free-field dimension 4 (c.f. (\ref{OPEa})). Therefore, the other two
operators, CPO and $O_{\bf 20}$, do not receive anomalous dimensions. To compute
$\Delta_{{\cal K}_{\bf 20}}^{(1)}$ one needs to know the free-field value of the
ratio $\frac{C_{OO{\cal K}_{\bf 20}}^2}{C_{{\cal K}_{20}}}$. This can be found by
considering, e.g., the $\tilde{\lambda}$-independent $vY^2$ terms in (\ref{p20}). By
using the CPWA of the scalar with dimension 2 and comparing $vY^2$ terms in
(\ref{OPEa}) with the ones in (\ref{p20}) one gets
$$
\frac{C_{OOO}^2}{3C_O}+\frac{C_{OO{\cal K}_{\bf 20}}^2}{4C_{{\cal K}_{\bf
20}}}=\frac{20}{3N^2}.
$$
Since $\frac{C_{OOO}^2}{C_O}=\frac{40}{3N^2}$ \cite{AFP} 
one finds the following free-field
value 
\bea \label{frel1} \frac{C_{OO{\cal K}_{\bf 20}}^2}{C_{{\cal K}_{\bf
20}}}=\frac{80}{9N^2}. \eea 
Analogously, analysis of the $v^2$ terms in
(\ref{OPEa}) and in (\ref{p20}) produces the free-field relation \bea
\label{frel2} \frac{C_{OOO}^2}{6C_O}+\frac{C_{OOO_{\bf 20}}^2}{C_{O_{\bf 20}}}-
\frac{C_{OO{\cal K}_{\bf 20}}^2}{4C_{{\cal K}_{\bf 20}}} =2+\frac{2}{3N^2} \eea
that further gives
$$
\frac{C_{OOO_{\bf 20}}^2}{C_{O_{\bf 20}}}=2+\frac{2}{3N^2}.
$$
Note that the same answer was found by studying the 4-point
function at strong coupling \cite{AFP}, that agrees with the conjectured
non-renormalization of the operator ${O_{\bf 20}}$.

Now the $\log{v}$-dependent term allows to find $\Delta_{{\cal K}_{\bf
20}}^{(1)}$: $\frac{C_{OO{\cal K}_{\bf 20}}^2}{8C_{{\cal K}_{\bf
20}}}\Delta_{{\cal K}_{\bf 20}}^{(1)}=\frac{10\tilde{\lambda}}{3N^2}$, {\it i.e.} ,
\bea
\label{adT20}
\Delta_{{\cal K}_{\bf 20}}^{(1)}=3\tilde{\lambda}.
\eea
Thus, the anomalous
dimension of ${\cal K}_{\bf 20}$ is the same
as the dimension of the Konishi field, hence
they are from the same multiplet.

Finally the $\tilde{\lambda}$-dependent terms without $\log{v}$ are due to the loop
correction to the free-field value of $C_{OO{\cal K}_{\bf 20}}$. 
Indeed, the $\tilde{\lambda} vY^2$ term in
(\ref{p20}) allows one to find
$C_{OO{\cal K}_{\bf 20}}^{(1)}=-3\tilde{\lambda}$. 
We can check the consistency of the assumption that there is
only one tensor operator in the ${\bf 20}$ which receives
corrections to its anomalous dimension and structure constant.
To this end we compute the term of order $v^2 Y$ by using the
found anomalous dimension and correction to the structure constant,
and see that it coincides with the corresponding term in (\ref{p20}).
\vskip 0.4cm
{\it 3.3 Projection in {\bf 84}}
\vskip 0.4cm

Projecting in ${\bf 84}$ we get for the leading
terms in the short-distance expansion
\bea \label{p84} &&\langle
O^{I_1}(x_1)O^{I_2}(x_2)O^{I_3}(x_3)O^{I_4}(x_4) \rangle |_{{\bf
84}} = \frac{C_{{\cal J}_{84}}^{I_1I_2}C_{{\cal
J}_{84}}^{I_3I_4}}{x_{12}^4x_{34}^4} \Biggl[
\frac{6\tilde{\lambda}}{N^2}v^2\left(1+Y\right)\log{v} \\
\nonumber && \hskip 1cm +
v^2\left(\left(2-\frac{2}{N^2}\right)(1+Y)-\frac{12\tilde{\lambda}}{N^2}
-\frac{9\tilde{\lambda}}{N^2}Y\right)
\Biggr]\, . \eea

A strong coupling result suggests that at finite $\lambda$ the
OPE of CPOs contains two operators
$O_{\bf 84}$ and ${\cal K}_{\bf 84}$
transforming in the irrep ${\bf 84}$.
The operator $O_{\bf
84}$ has protected both the dimension and the normalization
constants of the 2- and 3-point functions, while ${\cal K}_{\bf
84}$ is from the Konishi multiplet and receives anomalous
dimension.

At zeroth order in $\tilde{\lambda}$, the non-logarithmic term in
(\ref{p84}) gives for the free-field values of the normalization
constants\footnote{We exhibit a coefficient $1/N^2$ in front of 
$C_{{\cal K}_{\bf 84}}$ to emphasize the fact that  
a free-field operator $O^{fr}$ undergoes a splitting into a sum of
operators as   
$O^{fr}=O^{gr}+\frac{1}{N}O^{str}$, where $O^{gr}$ is dual to a
supergravity two-particle state and $O^{str}$ dual to a string state
\cite{AFP}. The same splitting applies to operators in irrep {\bf 175}.  
} 
$$
C_{O_{\bf 84}}+\frac{1}{N^2}C_{{\cal K}_{\bf 84}}=2-\frac{2}{N^2}\, .
$$
The constant $C_{O_{\bf 84}}$ is non-renormalized and is found
\cite{AFP} to be $C_{O_{\bf 84}}=2-\frac{6}{N^2}$. Therefore,
$C_{{\cal K}_{\bf 84}}=4$. The $\log{v}$ term
in (\ref{p84}) allows one to read off the anomalous dimension $\Delta_{{\cal
K}_{\bf 84}}^{(1)}$ of ${\cal K}_{\bf 84}$: $\Delta_{{\cal
K}_{\bf 84}}^{(1)}=3\tilde{\lambda}$, as it should be for the member of
the Konishi multiplet.
Finally, from the 
$\tilde{\lambda} v^2$ term in (\ref{p84}) we can find 
a correction $C_{OO{\cal K}_{\bf 84}}^{(1)}=-3\tilde{\lambda}$. 

\vskip 0.4cm
{\it 3.4 Projection in {\bf 105}}
\vskip 0.4cm

For the leading terms of the projection of the 4-point function
in ${\bf 105}$ we find
\bea \label{p105} &&\langle
O^{I_1}(x_1)O^{I_2}(x_2)O^{I_3}(x_3)O^{I_4}(x_4) \rangle |_{{\bf
105}} = \frac{C_{{\cal J}_{105}}^{I_1I_2}C_{{\cal
J}_{105}}^{I_3I_4}}{x_{12}^4x_{34}^4} \Biggl[
\frac{2\tilde{\lambda}}{N^2}v^3\left(1+\frac{3}{2}Y\right)\log{v} \\
\nonumber && \hskip 1cm + v^2\left(2+\frac{4}{N^2}\right) (1+Y) \Biggr]\, . \eea
The last formula shows that the first $\log{v}$-term appears at order $v^3$.
Therefore, all symmetric traceless rank-2k tensor operators of dimension $4+2k$
transforming in the ${\bf 105}$ have protected conformal dimensions, the lowest
operator among them is the double-trace operator $O_{\bf 105}$. The 
$\log{v}$-term in (\ref{p105})
indicates the appearance of the anomalous dimensions for the symmetric traceless
rank-$2k$ tensors of the canonical dimensions $6+2k$. However, at strong coupling
the first $\log{v}$-term appears only at order $v^4$ (see (4.16) of \cite{AFP}).
Thus, a free-field tensor operator of dimension $6+2k$ undergoes a splitting into
two operators, one has a protected dimension and normalization constants, another
one receives perturbatively an anomalous dimension and disappears at
strong coupling.

\vskip 0.4cm
{\it 3.5 Projection in {\bf 15}}
\vskip 0.4cm

Here we comment briefly on the projection in the irrep ${\bf 15}$
whose 
leading terms have the form \bea \langle
O^{I_1}(x_1)O^{I_2}(x_2)O^{I_3}(x_3)O^{I_4}(x_4) \rangle |_{{\bf 15}} =
\frac{C_{{\cal J}_{15}}^{I_1I_2}C_{{\cal J}_{15}}^{I_3I_4}}{x_{12}^4x_{34}^4}
\Biggl[\frac{8}{N^2}vY-\frac{16\tilde{\lambda}}{N^2}vY +\frac{8\tilde{\lambda}}{N^2}v Y\log{v}
\Biggr]\, . \label{p15} \eea 
The presence on the term $vY\log{v}$ shows the
appearance of the anomalous dimension for the vector operator ${\cal
K}_{\mu}^{{\cal J}_{15}}$ of dimension 3. At strong coupling, however, the term
$vY\log{v}$ is absent and the dimension 3 operator which is the
$R$-symmetry current $R_{\mu}^{{\cal J}_{15}}$ has protected conformal
dimension. 
Thus, at finite $\lambda$ the contribution to ${\bf 15}$ comes from two
operators, ${\cal K}_{\mu}^{{\cal J}_{15}}$ and $R_{\mu}^{{\cal J}_{15}}$. From
the $\tilde{\lambda}$-independent term $vY$ in (\ref{p15}) we read off the relation for
the free-field values of ratios of the normalization constants of these
operators: \bea \frac{C_{OO{\cal K}_{\mu}}^2}{2C_{\cal K_{\mu}}}+
\frac{C_{OOR_{\mu}}^2}{2C_{R_{\mu}}} =\frac{8}{N^2}\, .\label{relK15} \eea Taking
into account that $\frac{C_{OOR_{\mu}}^2}{2C_{R_{\mu}}} =\frac{8}{3N^2}$ we find
\bea \label{couplK15} \frac{C_{OO{\cal K}_{\mu}}^2}{2C_{{\cal K}_{\mu}}}=\frac{16}{3N^2}\, .
\eea Finally, from the $vY\log{v}$ term one obtains the anomalous dimension for
${\cal K}_{\mu}^{{\cal J}_{15}}$: \bea \label{adK15} \Delta_{{\cal
K}_{\mu}}^{(1)}=3\tilde{\lambda}, \eea {\it i.e.} the vector operator ${\cal
K}_{\mu}^{{\cal J}_{15}}$ is in the Konishi multiplet.

\vskip 0.4cm
{\it 3.6 Projection in {\bf 175}}
\vskip 0.4cm

Finally, projecting the 4-point function on irrep ${\bf 175}$ one obtains for the
leading terms the following expression: \bea \langle
O^{I_1}(x_1)O^{I_2}(x_2)O^{I_3}(x_3)O^{I_4}(x_4) \rangle |_{{\bf 175}} =
\frac{C_{{\cal J}_{175}}^{I_1I_2}C_{{\cal J}_{175}}^{I_3I_4}}{x_{12}^4x_{34}^4}
\Biggl[2v^2Y-\frac{4\tilde{\lambda}}{N^2}v^2Y +\frac{2\tilde{\lambda}}{N^2}v^2Y\log{v} \Biggr]\,
. \label{p175} \eea At strong coupling the first $\log{v}$ term occurs at order
$v^3Y$ ({\it c.f.} Section 4.6 of \cite{AFP}), while here it appears at order
$v^2Y$. Thus, at finite $\lambda$ the contribution of the lowest dimension operators
to the irrep.
{\bf 175} comes from two operators ${\cal K}_{\bf 175 }$ and $O_{\bf 175}$,
both with approximate dimension $5$. The first operator receives infinite
anomalous dimension at strong coupling, while the second one is non-renormalized
due to the shortening condition \cite{AFer2}. The $v^2Y$ term in (\ref{p175}) produces
for the free-field constants the following relation 
\bea
\label{relK175}
\frac{1}{N^2}C_{OO{\cal K}_{\bf 175}}+C_{OOO_{\bf 175}}=4 \, .
\eea
As was found in  \cite{AFP}
$C_{OOO_{\bf 175}}=4-\frac{8}{3N^2}$ and, therefore,
\bea
\label{couplK175}
C_{OO{\cal K}_{\bf 175}}=\frac{8}{3}\, .
\eea
Then the $\log{v}$ term allows one to find $\Delta_{{\cal
K }_{\bf 175}}^{(1)}=3\tilde{\lambda}$ justifying thereby that ${\cal K}_{\bf 175 }$ belongs
to the Konishi multiplet.

\section{Instanton contribution}
To analyze the instanton contribution to the 4-point function of the
lowest weight CPOs we use the results of
\cite{BGKR,DKMV,DHKMV1,DHKMV2}.  Firstly, we follow \cite{EPSS} to
write the 4-point function of the CPOs (2) as 
\bea \langle
\phi^{i_1j_1}(x_1)\phi^{i_2j_2}(x_2)\phi^{i_3j_3}(x_3) \phi^{i_4j_4}(x_4)\rangle
&=&
a_1(s,t)\frac{\d^{i_2j_2}_{\{i_1j_1\}}\d^{i_4j_4}_{\{i_3j_3\}}}{x_{12}^4x_{34}^4}
\nonumber \\
\nonumber
&+&a_2(s,t)\frac{\d^{i_3j_3}_{\{i_1j_1\}}\d^{i_4j_4}_{\{i_2j_2\}}}{x_{13}^4x_{24}^4}
+a_3(s,t)\frac{\d^{i_4j_4}_{\{i_1j_1\}}\d^{i_3j_3}_{\{i_2j_2\}}}{x_{14}^4x_{23}^4} \\
\nonumber &+&b_1(s,t)\frac{\d_{\{i_1j_1\}\{i_2j_2\}}^{\{i_3\{j_4i_4\}
j_3\}}}{x_{13}^2x_{14}^2x_{23}^2x_{24}^2}
+b_2(s,t)\frac{\d_{\{i_1j_1\}\{i_3j_3\}}^{\{i_2\{j_4i_4\}
j_2\}}}{x_{12}^2x_{14}^2x_{23}^2x_{34}^2}
\\ \label{basic}
&+&b_3(s,t)\frac{\d_{\{i_1j_1\}\{i_4j_4\}}^{\{i_2\{j_3i_3\}
j_2\}}}{x_{12}^2x_{13}^2x_{24}^2x_{34}^2} \, , \eea 
where $\phi^{ij}=\frac{1}{2^{1/2}
  \tilde\lambda}{\rm tr}[\phi^i\phi^j-\frac{1}{6}\d^{ij}\phi^2]$, $i,j=1,2,..,6$
and \bea s=\frac{x_{12}^2x_{34}^2}{x_{13}^2x_{24}^2},~~~~ t=
\frac{x_{14}^2x_{23}^2}{x_{13}^2x_{24}^2} \, . \eea
The traces are over $SU(N)$ adjoint indices and the $SO(6)$
group-theoretic $\d$-factors in (\ref{basic}) are products of
Kroenecker $\d^{'s}$ (c.f. \cite{EPSS}).  

Superconformal invariance implies that (\ref{basic}) is actually
determined in terms of only {\it two} arbitrary functions (e.g. $a_1$ and 
$b_2$) of $s$ and $t$. This fact allow us to restore the instanton
contribution to the full 4-point
function (\ref{basic}) from the results of
\cite{BGKR,DKMV,DHKMV1,DHKMV2} as follows. In the ${\cal N}=2$
formulation of ${\cal N}=4$ the six scalars in the fundamental of  
$SO(6)$ are decomposed in one complex scalar $\varphi$ and four
scalars comprising the
${\cal N}=2$ matter hypermultiplet. The complex scalar is
$\varphi=\phi^5+i\phi^6$ and one defines 
 \bea
 \mbox{tr}(\varphi^2)=\mbox{tr}(\phi^{55})-\mbox{tr}(\phi^{66})+2i
 \mbox{tr}(\phi^{56})=Y^{ij}\mbox{tr}(\phi^i\phi^j)\,, \eea
where $Y^{ij}=\d^{i5}\d^{j5}-\d^{i6}\d^{j6}+{\rm
  i}(\d^{i5}\d^{j6}+\d^{i6}\d^{j5})$. 
Then, from (\ref{basic}) using the nilpotency of $Y^{ij}$ we obtain
\bea \nonumber \langle
\mbox{tr}(\varphi^2)(x_1)\mbox{tr}(\varphi^2)(x_2)\mbox{tr}(\bar{\varphi}^2) 
(x_3) \mbox{tr}(\bar{\varphi}^2)(x_4)\rangle &=&
a_2(s,t)\frac{16}{x_{13}^4x_{24}^4}
+a_3(s,t)\frac{16}{x_{14}^4x_{23}^4} \\
&+&b_1(s,t)\frac{16}{x_{13}^2x_{14}^2x_{23}^2x_{24}^2}\, .\label{basic1} 
\eea 
The last
correlator is precisely the one computed in
\cite{BGKR,DKMV,DHKMV1,DHKMV2} and the result reads (omitting the
anti-instanton contributions)
\be \nonumber \langle
\mbox{tr}(\varphi^2)(x_1)\mbox{tr}(\varphi^2)(x_2)\mbox{tr}(\bar{\varphi}^2) 
(x_3) \mbox{tr}(\bar{\varphi}^2)(x_4)\rangle =16 Q\,x_{12}^4x_{34}^4
D_{4444}(x_1,x_2,x_3,x_4)\,,\label{inst} 
\ee
where the $D$-functions are defined in \cite{AFP} and we have absorbed
all the normalization factors into $Q$ defined as 
\be \label{Q}
Q=\frac{1}{4\tilde{\lambda}^4}\cdot
\frac{\sqrt{N}g^8_{YM}}{2^{33}\pi^{27/2}}k^{1/2}e^{2\pi i
k\tau}\sum_{d|k}\frac{1}{d^2}\cdot\frac{2^{30}\cdot 3^4}{16}\, ,\ee
with $\tau$ the usual complex Yang-Mills coupling.
The first factor in (\ref{Q}) is due to the normalization  of the
CPOs, the second factor comes from the $k$-instanton measure in the
large-$N$ limit and also takes into account the $R$-weight of the CPOs 
and the last one is the result of the integration of the fermionic zero modes. 
$Q$ is not a modular
invariant function, it is only the leading term of the modular invariant
expression in the large $g_{YM}$-limit.

In order now to read off the functions $a_1$, $a_3$ and $b_2$ in
(\ref{basic1}) from (\ref{inst}) we can exploit the
result of \cite{EPSS} according to which the above three
functions are in fact expressed in terms of only one function ${\cal
  F}(s,t)$ as 
\be \label{F}
b_1(s,t)=(s-t-1){\cal F}(s,t) \, ,\,\,\, a_2(s,t)= {\cal F}(s,t)
\, ,\,\,\, a_3(s,t)= t{\cal F}(s,t) \, . 
\ee 
The function ${\cal F}(s,t)$ should satisfy the following
crossing-symmetry properties
\bea \label{rel}
{\cal F}(s,t)={\cal
F}(t,s)=\frac{1}{t}{\cal F}(s/t,1/t) \,.
\eea
From (\ref{basic1}) and (\ref{F}) we then obtain
\bea
{\cal
F}(s,t)=Q\frac{s}{t^3}D(s,t)=Q\frac{v^3}{u^2}\bar{D}_{4444}(v,Y)\, ,
\eea
where
\bea
\nonumber
D(s,t) & =& \bar{D}_{4444}(s,t)\nonumber \\
&=& 2K\int \rmd t_1...\rmd t_4(t_1t_2t_3t_4)^3exp\Biggl[
-t_1\left( t_2+t_3+t_4\right)
-t_2t_3-\frac{1}{t}t_2t_4-\frac{s}{t}t_3t_4 \Biggr], 
\eea
and $K$ was defined in \cite{AFP}. 
Using the above  integral representation it is easy to check 
that the function ${\cal F}(s,t)$
does satisfy the relations (\ref{rel}). 

Multiplying (\ref{basic}) with $C_{ij}^I$ we find the instanton contribution to
the {\it complete} 4-point function of the lowest weight CPOs as
\ba
\label{inst4pt}
&&\langle
O^{I_1}(x_1) O^{I_2}(x_2)O^{I_3}(x_3)O^{I_4}(x_4)\rangle
|_{inst}=\frac{1}{x_{12}^4x_{34}^4}\Biggl[ \d^{I_1I_2}\d^{I_3I_4}A_1(v,Y)
+\d^{I_1I_3}\d^{I_2I_4}A_2(v,Y)\nonumber \\
&&\hspace{-0.5cm}+\d^{I_1I_4}\d^{I_2I_3}A_3(v,Y)
+C^{I_1I_2I_3I_4}B_2(v,Y)+C^{I_1I_3I_2I_4}B_1(v,Y)+C^{I_1I_3I_4I_2}B_3(v,Y)
\Biggr]\, ,\ea where
\bea 
\nonumber
A_1(v,Y)&=&a_1(s,t)=Q\frac{v^3}{u}\bar{D}_{4444}(v,Y) \, ,\\
\nonumber
A_2(v,Y)&=&u^2a_2(s,t)=Q v^3\bar{D}_{4444}(v,Y) \, ,\\
\nonumber
A_3(v,Y)&=&v^2a_3(s,t)=Q\frac{v^4}{u}\bar{D}_{4444}(v,Y) \, ,\\
B_1(v,Y)&=&uvb_1(s,t)=Q\left(u-\frac{u}{v}-1\right)\frac{v^4}{u}\bar{D}_{4444}(v,Y)\, ,\\
\nonumber
B_2(v,Y)&=&vb_2(s,t)=Q\left(1-u-\frac{u}{v}\right)\frac{v^4}{u^2}\bar{D}_{4444}(v,Y)\, ,\\
\nonumber
B_3(v,Y)&=&ub_3(s,t)=Q\left(\frac{u}{v}-u-1\right)\frac{v^3}{u}\bar{D}_{4444}(v,Y)
\, . \eea
The $\bar{D}_{4444}$ function has the following decomposition
\bea
\bar{D}_{4444}&=&\frac{5\pi^2}{108}\sum_{n,m=0}^{\infty}
\frac{Y^m}{m!}\frac{v^n}{(n!)^2}\frac{\G^2(n+4)\G^2(n+m+4)}{\G(8+2n+m)}
\\
\nonumber
&\times& \left[-\log v+2\psi(n+1)-2\psi(n+4)-2\psi(n+m+4)+2\psi(8+2n+m)
\right] \, .
\eea

We are now ready to analyze the contribution of the instantons to OPE
of the lowest weight CPOs. Firstly we consider the short-distance expansion for the
projection in the singlet. We find that the leading terms
are given by
\bea
\nonumber \langle
O^{I_1}(x_1)O^{I_2}(x_2)O^{I_3}(x_3)O^{I_4}(x_4) \rangle |_{{\bf 1
}} &=& \frac{\pi^2Q~~\d^{I_1I_2}\d^{I_3I_4}}{x_{12}^4x_{34}^4}\Biggl[
-\frac{1}{84}(1+Y)v^2\log v  \\
\label{inst1}
&-&v^2\left(\frac{451}{17640}+\frac{139}{4410}Y\right)
\Biggr] \, .
\eea
This clearly shows that the leading contribution to the OPE comes from 
a scalar operator
of approximate dimension 4. Our experience at strong coupling teaches
us that the only such operator is the double-trace operator $O_1$ discussed in
Section 3.   
On the other hand, due to the absence of $v\log v$
and $vY^2\log v$-terms, the contribution of the Konishi fields ${\cal K}$
and ${\cal K}_1$, and of the operator $\Xi_{\mu\nu}$ 
are absent. We conclude that the
Konishi multiplet as well as the multiplet built on $\Xi_{\mu\nu}$ 
receive only perturbative but not instanton
corrections.\footnote{The absence of the instanton corrections   
to the Konishi multiplet was already noted in \cite{BKRS1}.} Furthermore, we observe
in (\ref{inst1}) the absence of the contribution of the stress tensor, 
in agreement with the known non-renormalization theorem for this
operator.  

For the projection on {\bf 20} the leading contribution reads as
\bea
\nonumber \langle
O^{I_1}(x_1)O^{I_2}(x_2)O^{I_3}(x_3)O^{I_4}(x_4) \rangle |_{{\bf 20
}} &=& \frac{\pi^2Q~~C_{{\cal J}_{20}}^{I_1I_2}C_{{\cal
J}_{20}}^{I_3I_4}}{x_{12}^4x_{34}^4}\Biggl[
-\frac{5}{252}v^2Y^2\log v
\\ \label{inst20}
&-&\frac{451}{10584}v^2\left(Y^2-v\right)
\Biggr] \,
\eea
and comes from a second rank tensor
of the canonical dimension 6. The contribution from the
Konishi multiplet is again absent. Recall that at strong coupling
we have found that the first operators receiving anomalous dimension
are scalar and tensor operators of approximate dimension $6$ which we
therefore identify with the operators appearing in (\ref{inst20}).

These two examples at hand, i.e. projection in the singlet and in
${\bf 20}$, allow us to make a 
general observation: the (double-trace) operators receiving finite anomalous
dimensions at strong coupling also receive instanton contributions.
The instanton contribution to the (single-trace) operators
with infinite anomalous dimensions at strong coupling is absent.

Let us examine the other irreps. One gets the following leading behavior \\
for {\bf 84}:
\bea
\nonumber \langle
O^{I_1}(x_1)O^{I_2}(x_2)O^{I_3}(x_3)O^{I_4}(x_4) \rangle |_{{\bf 84
}} &=& \frac{\pi^2Q~~C_{{\cal J}_{84}}^{I_1I_2}C_{{\cal
J}_{84}}^{I_3I_4}}{x_{12}^4x_{34}^4}\Biggl[
-\frac{1}{28}v^3\log v -\frac{451}{5880}v^3
\Biggr] \, ;
\eea
for {\bf 105}:
\bea
\nonumber \langle
O^{I_1}(x_1)O^{I_2}(x_2)O^{I_3}(x_3)O^{I_4}(x_4) \rangle |_{{\bf 105
}} &=& \frac{\pi^2Q~~C_{{\cal J}_{105}}^{I_1I_2}C_{{\cal
J}_{105}}^{I_3I_4}}{x_{12}^4x_{34}^4}\Biggl[
-\frac{1}{84}v^4\log v -\frac{451}{17640}v^4
\Biggr] \, ;
\eea
for {\bf 15}:
\bea
\nonumber \langle
O^{I_1}(x_1)O^{I_2}(x_2)O^{I_3}(x_3)O^{I_4}(x_4) \rangle |_{{\bf 15
}} &=& \frac{\pi^2Q~~C_{{\cal J}_{15}}^{I_1I_2}C_{{\cal
J}_{15}}^{I_3I_4}}{x_{12}^4x_{34}^4}\Biggl[
-\frac{1}{21}v^2Y\log v -\frac{451}{4410}v^2Y
\Biggr] \, ;
\eea
for {\bf 175}:
\bea
\nonumber \langle
O^{I_1}(x_1)O^{I_2}(x_2)O^{I_3}(x_3)O^{I_4}(x_4) \rangle |_{{\bf 175
}} &=& \frac{\pi^2Q~~C_{{\cal J}_{175}}^{I_1I_2}C_{{\cal
J}_{175}}^{I_3I_4}}{x_{12}^4x_{34}^4}\Biggl[
-\frac{1}{84}v^3Y\log v -\frac{451}{17640}v^3Y
\Biggr] \, .
\eea
Comparison of these formulae with the analogous strong coupling results
confirms the above observation. Moreover, we see that the instanton 
contributions do not spoil the non-renormalization property of certain
towers of double-trace operators found in \cite{AFP}. 
Indeed, the scalar operator $O_{\bf 20}$
in {\bf 20}, all the rank $2k$ tensors of dimension $4+2k$ in ${\bf 84}$
and of dimension $4+2k$, $6+2k$ in {\bf 105}, and all the rank $2k+1$
tensors of dimension $5+2k$ in ${\bf 175}$ are non-renormalized in the 
instanton background.  

The absence in the instanton OPE of 
corrections to the operators in the Konishi multiplet can be
easily understood at follows. Instanton corrections
to the normalization constant $C_{OO{\cal K}}$ and to the anomalous dimension 
of $\cal K$ are encoded into the 3- and 2-point functions of ${\cal K}$ in
the instanton background.  
According to the prescription of  \cite{BGKR}-\cite{DHKMV2}, to
calculate the
correlation functions of composite operators in the instanton background we 
replace the latter by their instanton background expressions; the resulting
correlation function is then non-zero only if it contains all 16 fermionic zero
modes. This is needed in order to saturate the fermionic integration measure. It
is then easy to see that the following 3-point function 
\bea 
\langle OO{\cal K}\rangle \sim \langle {\mbox
tr}(\phi^{(i_1}\phi^{i_2)}){\mbox tr}(\phi^{(i_1}\phi^{i_2)}){\mbox
tr}(\phi^k\phi^k)\rangle \,,
\eea 
contains only 12 fermionic zero modes and therefore it should be zero
in the instanton background.
Such arguments can be generalized to the whole Konishi
multiplet. Indeed, as was shown in \cite{IS}, the number of the fermion zero modes
for an operators $O^{(q)}$ from the Konishi multiplet is $4-|q|$, where $q$ is a
$U(1)_Y$ charge. The 3-point function of $O^{(q)}$ with two lowest-weight CPOs
provides only $12-|q|$ zero modes and, therefore, vanishes.

Application of the same arguments to the double-trace operator  ${\cal O}_{\bf
20}$ shows that it can, in principle, receive instanton corrections.
However, our explicit OPE calculation shows that it is not the
case. This means that the particular dynamics that keeps ${\cal O}_{\bf
20}$ non-renormalized is not affected by instantons. 

Concerning instanton corrections to the other multiplets we found,
their existence 
is in agreement with the general considerations of \cite{IS}. An operator $O^{(q)}$
from a generic long multiplet provides $8-|q|$ zero modes, so that the 2- and
3-point correlation functions involving $O^{(q)}$ are non-zero only if $q=0$.
Thus, in our OPE the leading operators that receive instanton corrections and
survive at strong coupling with finite anomalous dimensions, are either primary
operators (or conformal descendants of the primary operators), $O_p$
generating long multiplets or have the form $Q^k\bar{Q}^kO_p$.

Coming back to the singlet projection we now compute the instanton
contribution $\D^{(i)}$ to the anomalous dimension of $O_1$. If we assume that the small
parameter at hand is $\sim N^{1/2}/N^4$, then
\bea
\frac{C_{OOO_1}^2}{2C_{O_1}}\D^{(i)}_{{\bf 1}}=-\frac{\pi^2Q}{84}.
\eea
Substituting for $\frac{C_{OOO_1}}{C_{O_1}}$ its free-field value $1/10$
we get the anomalous dimension
\bea
\D^{(i)}_{{\bf 1}}=-\frac{5\pi^2Q}{21} \, .
\eea

By using the results of \cite{AFP} we 
have also checked that the same ratio occurs for the leading operators 
in {\bf 84}, {\bf 105} and {\bf 175} indicating thereby a universal behavior
\bea
\label{equal}
\frac{\D^{(i)}_{{\bf 1}}}{\D^{(s)}_{{\bf 1}}}=
\frac{\D^{(i)}_{{\bf 15}}}{\D^{(s)}_{{\bf 15}}}=
\frac{\D^{(i)}_{{\bf 84}}}{\D^{(s)}_{{\bf 84}}}=
\frac{\D^{(i)}_{{\bf 105}}}{\D^{(s)}_{{\bf 105}}}=
\frac{\D^{(i)}_{{\bf 175}}}{\D^{(s)}_{{\bf 175}}}=\frac{5\pi^2}{336}QN^2 \, . 
\eea

\section{Concluding remarks}

In this work we have extended the OPE analysis of the lowest weight CPOs
initiated in \cite{AFP}, to include 2-loop and instanton
contributions. At the perturbative level, we found that it is not sufficient to simply
deform the free-field operator algebra by allowing for anomalous
dimensions and corrections to the coupling in order to account for
the 2-loop corrections. What is required  is a splitting of 
various free-field operators into operators 
belonging to distinct  supermultiplets
which behave in a different way under the RG-flow. We have explicitly
demonstrated this splitting in the case of the stress tensor and the
$R$-current of the theory. Our results are consistent with earlier
calculations by Anselmi in \cite{An,An1}. Furthermore, we argued that
a similar splitting occurs also for free-field theory double-trace
operators, e.g., they split into supermultiplets which behave in
perturbation theory either like the Konishi multiplet or acquire
anomalous dimensions $\sim \tilde{\lambda}/N^2$. The latter multiplets
are the ones which survive at strong coupling and get non-zero
anomalous dimensions. The above splitting seems to be necessary in
order to explain the fact that while at any order in perturbation
theory we expect a mixing of many operators with the same free-field
dimension, at strong coupling only one of the operators is present  while
all the others decouple. Nevertheless, an explicit calculation of the 
the 2-loop anomalous dimension of  the split operators would require
the knowledge of 4-point functions including operators other than the
lowest weight 
CPOs, e.g., the 4-point function of Konishi scalars. We believe that
this is an interesting project.

We also found that the  instantons give contributions only to operators
which acquire non-zero and finite anomalous dimensions at
strong coupling. In particular, instantons do not contribute neither
to protected nor to operators in the Konishi multiplet. This is
consistent with general arguments given in \cite{IS} concerning the
vanishing of the three-point functions of operators with non-zero
$U(1)_Y$-charge in the instanton background. 

The fact that instantons contribute only to operators which correspond
to ``two-particle'' modes of classical supergravity,
points to an interesting connection between the latter modes and
D-particles modes which is worth further study. Intuitively, the
corrections to the anomalous dimensions of the supergravity modes
come from a corresponding change of their energy in the presence of
$D$-particles. On the other hand,
it can be seen that the universal behavior (\ref{equal}) for the
leading operators in the OPE is a consequence of the fact that 
the correlation function of CPOs is defined (up to the free-field contribution) by 
a unique function ${\cal F}(v,Y)$. If one projects in a given irrep ${\cal J}$,
one gets 
\bea
\nonumber \langle
O^{I_1}(x_1)O^{I_2}(x_2)O^{I_3}(x_3)O^{I_4}(x_4) \rangle |_{{\cal J}} &=& 
\frac{C_{{\cal J}}^{I_1I_2}C_{{\cal
J}}^{I_3I_4}}{x_{12}^4x_{34}^4}\Biggl[ h_{\cal J}(v,Y)+f_{\cal J}(v,Y){\cal F}(v,Y)\Biggr]\, ,
\eea
where $h_{\cal J}(v,Y)$ is a free-field contribution, and 
$f_{\cal J}(v,Y)$ is some function depending on the irrep we consider. 
Now we see that if the strong coupling and instanton 4-point functions 
are described by ${\cal F}^{(s)}(v,Y)$ and ${\cal F}^{(i)}(v,Y)$ respectively, 
then after subtracting the free-field theory contribution, 
the ratio of the strong to 
the instanton contributions will be one and the same for all irreps
since the function $f_{\cal J}(v,Y)$
does not depend on the regime we consider and, therefore, cancels out.
What is more important is that the short-distance 
expansion of the {\it both} 
functions $f_{\cal J}(v,Y){\cal F}^{(i,s)}(v,Y)$ starts as
$f_{\cal J}(v,Y){\cal F}^{(i,s)}(v,Y)=v^{\frac{\D-l}{2}}G_l^{(i,s)}(Y)\log v+...$, 
where the function $G_l^{i,s}(Y)$ may be represented as 
$G_l^{(i,s)}(Y)=a_{\cal J}(Y)f^{(i,s)}(Y)=aY^l+...$ and it encodes the leading 
contribution of the rank-$l$ tensors of 
a canonical dimension $\D$. Therefore, 
the equalities (\ref{equal}) stem from the fact 
that the ratio of the instantons to strong-coupling corrections 
$\frac{G_l^{(i)}(Y)}{G_l^{(s)}(Y)}$ does not depend on an irrep we consider. 


Finally, we wish to comment on the fact that we have confirmed both in
perturbation theory and in the instanton background the
non-renormalization of various operators, most importantly the
scalar double-trace operator in the ${\bf 20}$ with dimension
4. The dimension of the latter operator is not protected by unitarity
constraints and therefore it is subject to a genuine dynamical
protection. Being a double-trace operator, it is difficult to
find the corresponding supergravity mode. Furthermore, since it is a
marginal operator we may use it to deform the ${\cal N}=4$ SYM$_4$ action 
preserving the conformal invariance to leading order in the deformation.
It would be of great interest to find out if this operator is exactly marginal
as in this case it defines a particular RG flow (fixed line), that  
might lead to a new non-trivial ${\cal N}=0$ CFT.

\vskip 0.3cm
{\bf Acknowledgements}
We would like to thank A. Tseytlin for useful comments on the manuscript. 
G. A. and T. P. are grateful to S. Ferrara and P. Fre' 
for useful discussions. G. A. would like to acknowledge helpful
conversations with S. Theisen.
The work of G. A. was supported by the DFG and by the European Commission 
RTN programme HPRN-CT-2000-00131 in which G. A. is associated to U. Bonn, and 
in part by RFBI grant N99-01-00166. 
The work of S.F. 
was supported by the U.S. Department of Energy under grant
No. DE-FG02-91ER-40690 and in part by RFBI grant N99-01-00190. 
T. P. was supported by the E.U. under the
program RTN1-1999-00116.

\end{document}